\documentclass[conference]{IEEEtran}
\makeatletter
\def\ps@headings{%
\def\@oddhead{\mbox{}\scriptsize\rightmark \hfil \thepage}%
\def\@evenhead{\scriptsize\thepage \hfil \leftmark\mbox{}}%
\def\@oddfoot{}%
\def\@evenfoot{}}
\makeatother
\usepackage{tabularx}
\usepackage{booktabs}
\usepackage{fixltx2e}

\usepackage{graphicx}
\usepackage{amssymb}
\usepackage[cmex10]{amsmath}
\usepackage{mdwmath}
\usepackage{mdwtab}
\usepackage{eqparbox}
\usepackage{algorithm}
\usepackage{algpseudocode}

\usepackage{caption}
\usepackage{subcaption}

\begin{document}

\title{Analysis of urban traffic data sets for VANETs simulations}

\author{\IEEEauthorblockN{Alice Castellano\IEEEauthorrefmark{1} and
Francesca Cuomo\IEEEauthorrefmark{4}}
\IEEEauthorblockA{\IEEEauthorrefmark{1}University of Roma ``Sapienza'', Rome, Italy Email: alice.alice@alice.it}
\IEEEauthorblockA{\IEEEauthorrefmark{4}DIET at University of Roma ``Sapienza'', Rome, Italy Email: francesca.cuomo@diet.uniroma1.it}}

\maketitle


\section{Introduction}
\label{sec:intro}
Vehicular Ad-hoc Networks (VANET) are self-organized, distributed communication networks built up from moving vehicles   where each node is characterized by variable speed, strict limits of freedom in movement patterns and a variety of traffic dynamics.
In the last few years vehicular traffic is attracting a growing attention from both  industry and research, due to the importance of the related applications, ranging from traffic control to road safety. However, less attention has been paid to the modeling of realistic user mobility through real empirical data that would allow to develop much more effective communication and networking schemes.
The aim of this study is then twofold: on one hand we derived real mobility patters to be used in a VANET simulator and on the other hand we simulated the VANET data dissemination achieved with different broadcast protocols in our real traffic setting.

\section{Analysis of GPS data traces}
\label{sec:analysis}
In order to achieve a realistic modeling for VANETs, we analyzed  1.17 GB of raw data in Microsoft Excel sheet format containing  GPS traces from  more than  50.220 vehicles.
These GPS traces \cite{1} refer to  $2\%$  of total vehicles during the month of  May 2010  measured in the  Grande Raccordo Anulare (GRA) area, the circular  highway surrounding the city of Rome.
Each vehicle sends its GPS trace record every 30 seconds. A variety of information was captured in each record, including the record ID, vehicle ID, geographical coordinates, speed, quality of GPS signal.
The first step of our analysis was based on the segmentation of the GRA area in 29 different segments of length $L_j$, $j=1,...,29$, where the main exits on the GRA highway are the starting and ending points of each segment.
Then vehicles were divided in two sets according to their traffic direction : clockwise and counterclockwise. This last step allowed us to calculate the distance of each vehicle from the closest preceding one.
Four different time periods of four hours each have been considered: 1.) [7a.m-11a.m.), 2.) [11a.m. -3p.m.), 3.) [3p.m. -7p.m.)  and 4.) [7p.m.-11p.m.).
Inter-vehicle distance distribution and  speed distribution were obtained  for each of the four time periods.
This analysis showed that the highest density of vehicles is in the time period between 3p.m. and 7p.m., which has the largest number of detected vehicles (9732 vehicles). In this range of time we  also have the largest number of vehicles with inter-distance $<=50\;m$ and the lowest average speed ($64.4\; km/h$) compared with the other three time periods considered.

\begin{figure}[t]
\begin{center}
\includegraphics[width=0.3 \textwidth]{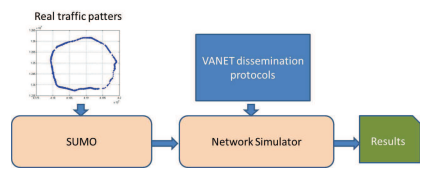}
\caption{Simulation framework.}
\label{fig:simulation}
\end{center}
\end{figure}

\begin{figure}[t]
\begin{center}
\includegraphics[width=0.3 \textwidth]{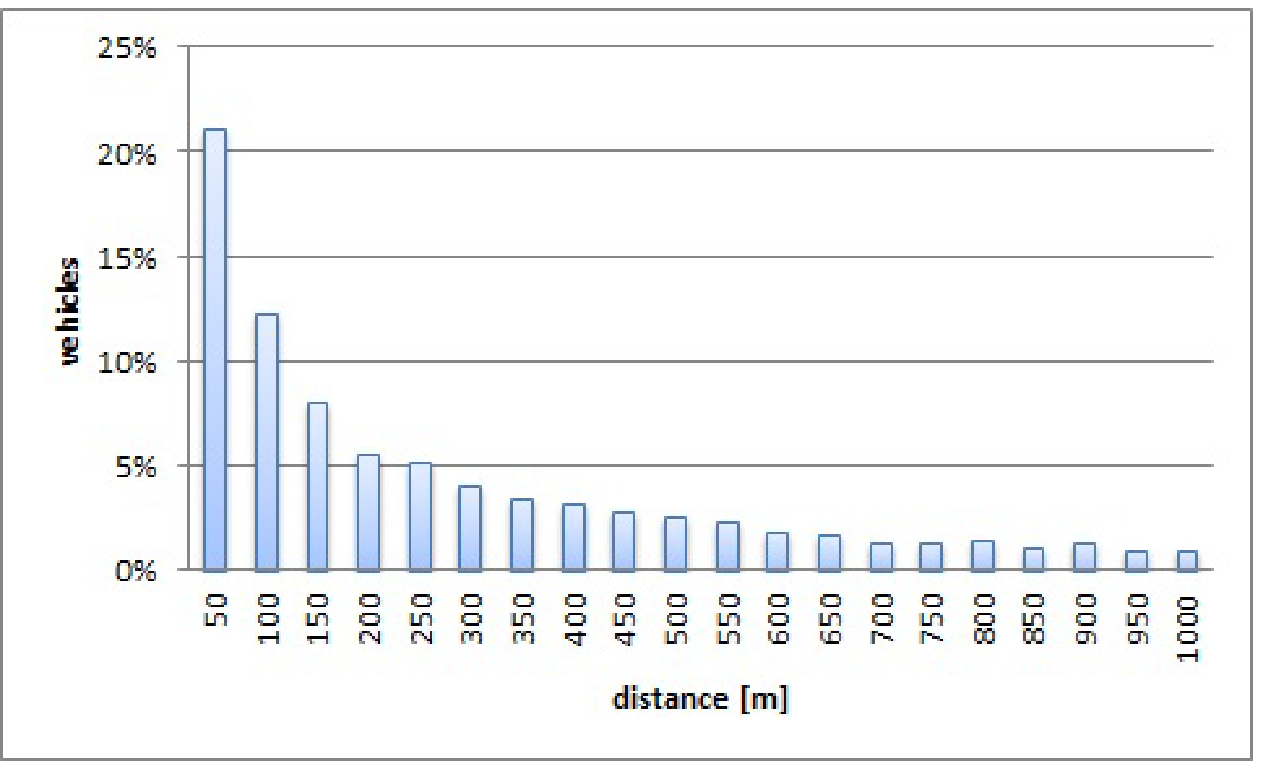}
\caption{Inter-vehicle distance in case of time period 3p.m. -7p.m.}
\label{fig:distance_1}
\end{center}
\end{figure}

\begin{figure}[t]
\begin{center}
\includegraphics[width=0.3 \textwidth]{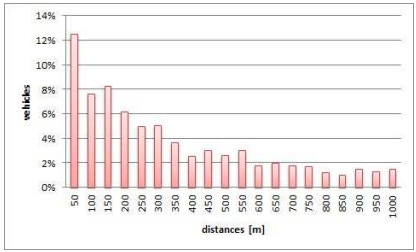}
\caption{Inter-vehicle distance in case of time period 7p.m.-11p.m.}
\label{fig:distance_2}
\end{center}
\end{figure}

On the contrary, as we expected, the sparsest time period  is the one between 7p.m. and 11p.m. with the lowest number of vehicles (978), largest inter-vehicle distances and highest average speed ($95.5\;km/h$).

\section{Injection of real data traces in the simulations and performance analysis}
\label{sec:perf}
To set up a realistic mobility simulation of the urban traffic in the GRA area the available data have been re-densified to have a realistic amount of vehicles ($100\%$).
The mathematical model \cite{2} used for densification  is briefly described by the following relations, where $h$ is the sampling temporal interval (30 sec in our study), $v_{i}$ is the average speed of vehicle $i$, $m_j$ is the number of the estimated traveling vehicles  in the $j$ segment, $n_j$ is the number of detected GPS signals in the segment $j$, $L_j$ is the length of the $j$ segment,  $a$ is the probe vehicles penetration rate ($2\%$ in our study) and $q_j$ is the estimated flow on the segment $j$.
We define $d_i=v_i*h$ and we have on the segment $j$:
\begin{equation}
m_j=\sum_{i=1}^{n_j} d_i/L_j
\end{equation}
and the resulting flow as $q_j=m_j/a$.

The above densification rules have been applied to the traffic flows in the time period with the highest density, 3-7p.m. and have been used to inject on each segment $j$ and amount of vehicles equal to $q_j$. It is to be notices that vehicles are injected and extracted in correspondence to each of the GRA exits.  Then the generated flows have moved in the GRA scenario by using SUMO (Simulation Of Urban Mobility) \cite{3} and then used to run communication protocols by using  NS-2 (Simulator Network 2) \cite{4} and MOVE (Mobility model generator for Vehicular Networks) \cite{5} (see Fig. \ref{fig:simulation} ). These tools provide, respectively, vehicles mobility, simulation of network protocols and a graphical interface between networking and mobility.
The protocols simulated to provide message spreading in the proposed scenario are:
\begin{itemize}
  \item Flooding: simply floods the network with messages.
  \item DBF: based on DDT (Distance Defer Transmission) \cite{6}, a node $A$  receiving a message from node $B$ waits $T_{max}(1-d_{AB}/R_{max})$  before forwarding the packet, where $R_{max}$ is an upper bound of the radio range and $d_{AB}$ is the Euclidean distance between $A$ and $B$.
  \item DBF hop count: enhanced version of DBF, avoids erroneous propagation interruptions that in standards DBF are noticed when two forwarders are so close that their timers are almost the same, so they forward the same message, which triggers the inibition rule at the next hop, thus stopping propagation.
  \item RND: a random selection of the forwarder where each node receiving a message sets a timer to a value between 0 and 100 ms before forwarding and when a forward takes palace this inhibits neighbor nodes.
\end{itemize}

\begin{figure}[tb]
        \centering
        \begin{subfigure}[tb]{0.3\textwidth}
                \centering
					\includegraphics[width=\textwidth]{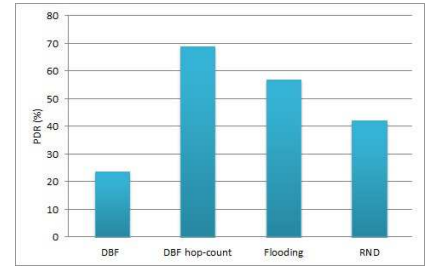}
                \caption{PDR of the tested protocols}
                \label{fig:PDR}
        \end{subfigure}
        \begin{subfigure}[tb]{0.32\textwidth}
                \centering
					\includegraphics[width=\textwidth]{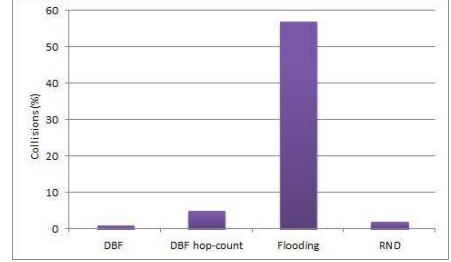}
                \caption{MAC collisions}
                \label{fig:COLL}
        \end{subfigure}
		\begin{subfigure}[tb]{0.32\textwidth}
                \centering
                \includegraphics[width=\textwidth]{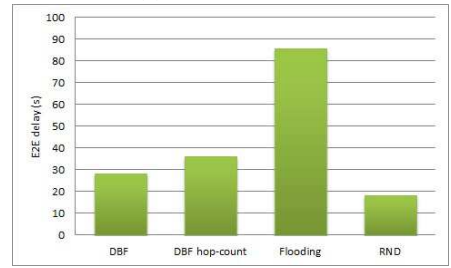}
                \caption{End to End average delay}
                \label{fig:E2E}
        \end{subfigure}
 \caption{Simulation results}
\label{fig:results}
\end{figure}

%
%
	
Simulations last 100s  and were performed with  a bit-rate of $160\;kbps$, placing a single RSU (Road Side Unit) near an Exit ($\#19$), on the southern part of the GRA highway.
The metrics used to compare  performances of the different protocols are: MAC Collisions, PDR (Packet Delivery Ratio, average fraction of messages received by a node), Average End-to-end delay (average time needed for a message to reach the farthest node from the RSU).
Flooding protocol achieves a good performance only when few nodes are present but behaves really bad in presence of more traffic. As we expected, it has the highest number of collisions.
DBF hop-count achieves the best results compared to the others, avoiding the packets to stop. The graphs in Fig. \ref{fig:results} show that DBF hop-count obtains the highest PDR value, low number of Collisions and an acceptable value of Average End-to-end delay.

\section*{Acknowledgments}
Authors are grateful to Mario De Felice for the work done in the NS-2 simulations set up.

{}	


\end{document}